\documentclass[letterpaper,12pt]{article}   
\usepackage{osajnl2} 

\begin{document}

\title{Inter-pixel crosstalk in Teledyne Imaging Sensors (TIS) H4RG-10 detectors}


\author{Rachel P. Dudik ,$^{1}$ Margaret E. Jordan,$^7$ Bryan N. Dorland,$^{1}$ Daniel Veillette,$^1$ Augustyn Waczynski,$^2$ 
Benjamin F. Lane,$^3$ Markus Loose,$^5$ Emily Kan,$^2$ James Waterman,$^4$ Chris Rollins,$^8$ and Steve Pravdo$^{6}$}
\address{$^1$United States Naval Observatory, \\ 3450 Massachusetts Avenue, NW,
Washington, D.C. 20392, USA}
\address{$^2$Goddard Space Flight Center, NASA \\ 8800 Greenbelt Road,
Greenbelt, MD 20771, USA}
\address{$^3$Charles Stark Draper Laboratory, Inc., \\ 555 Technology Sq.,
Cambridge, MA 02139, USA}
\address{$^4$Optical Sciences Division, Naval Research Laboratory, \\4555 Overlook Ave., SW ,
Washington, D.C. 20375, USA}
\address{$^5$Markury Scientific, Inc., \\ 518 Oakhampton Street,
Thousand Oaks, CA 91361, USA}
\address{$^6$Jet Propulsion Laboratory, NASA, \\ 4800 Oak Grove Drive,
Pasadena, CA 91109, USA}
\address{$^7$Computational Physics, Inc., \\ 8001 Braddock Rd.,
Springfield, VA 22151, USA}
\address{$^8$Research Support Instruments, Inc., \\4325-B Forbes Boulevard.
Lanham, MD 20706, USA}

\begin{abstract}

CMOS-hybrid arrays have become competitive optical detectors for use in ground- and space-based astronomy.  Inter-pixel capacitance is one source of error that appears in most CMOS arrays.  In this paper we use a single pixel reset method to model inter-pixel capacitance (IPC).  We combine this IPC model with a model for charge diffusion to estimate the total crosstalk on H4RG-10 arrays.  Finally, we compare our model results to $^{55}$Fe data obtained using an astrometric camera built to test the H4RG-10 B0 generation detectors.
\end{abstract}

                          

\section{Introduction}

Complementary Metal-Oxide Semiconductor (CMOS) sensors have become a competitive astronomical ground- and space-based detector solution.  CMOS sensors have a flexible readout structure that allows a single pixel or group of pixels on the array to be read out or reset at any time without disturbing or reading out the rest of the array.  This random-access, non-destructive read capability is ideal for dynamic range-driven astronomical applications, since it permits bright and faint objects to be observed simultaneously using a single detector.   In addition to the flexible readout, CMOS sensors are naturally less sensitive to radiation than more traditional detectors like Charge Coupled Devices (CCDs), since damage to one pixel in the array does not adversely affect subsequent pixels in a row or column of the array.  This inherent radiation hardness is particularly appealing for space-based applications.

While the readout capabilities of CMOS sensors are ideal for a variety of observing strategies, the fill factor of each pixel is significantly reduced because the readout circuitry is implanted directly on the photodetector material.  Janesick et al. \cite{Janesick2005} describe how the CMOS-hybrid focal planes have been developed to address this issue.  A CMOS-hybrid sensor is a CMOS device or Readout Integrated Circuit (ROIC), mated with a layer of photodetector material.  The two layers are typically joined together using indium bump bonds.  The resultant hybrid SCA is back-illuminated, and combines the flexible readout of the CMOS with a CCD-like fill factor of 100\%.   The United States Naval Observatory (USNO) has used CMOS-hybrid detectors to take advantage of this performance and flexibility, as part of its development of very large format focal plane technologies, in support of the Joint Milli-Arcsecond Pathfinder Survey (JMAPS) astrometry mission.

USNO has been testing large format, Teledyne Imaging Sensors (TIS) H4RG-10 Hybrid Visible Silicon Imager (HyViSI) Sensor Chip Assemblies (SCAs) since the development of the first generation-A1 detector in 2006, as described by Dorland et al. in 2007 \cite{Dorland2007}.  In 2008 USNO supported development of a second generation-A2 H4RG-10 with significantly lower dark current than the A1 predecessor, again described by Dorland et al. in 2009 \cite{Dorland2009}  A third generation detector, the H4RG-10 B0 detector, was fabricated in 2010 with JMAPS support to address yield and pixel operability requirements.  This third generation of detector has very low dark current and noise properties comparable to many CCDs.  The quantum efficiency is in excess of 80\% across multiple wavelengths and the non-linearity is $<1\%$. Additionally, with JMAPS risk reduction support, Hubbs et al. in 2011 \cite{Hubbs2011}, describes how TIS was able to increase pixel operability to better than 99.9\%.  Here, operability specifically refers to the percent of pixels that are fully connected without shorting.

While these flexible and low-noise CMOS hybrid arrays are excellent for most astronomical applications, the H4RG-10 show higher levels of crosstalk, or more specifically:  the Inter-Pixel Capacitance (IPC) component of crosstalk than CCDs.  We also note that some CMOS pixel circuit designs do not exhibit high levels of IPC.  However, the chosen design of the H4RG-10 has advantages that are difficult to achieve with these other designs, namely low read noise, low dark current and low power consumption. Indeed, most low background applications prefer the source-follower approach despite the IPC problem (eg. JMAPS H4RG-10 B0), while higher background applications typically use one of the other design options.  

As discussed in detail below, crosstalk can be problematic for applications like astrometry, because it has the effect of blurring the point spread function (PSF) of the photon source, resulting in lower effective signal to noise (S/N) in each pixel and high centroiding errors (See also \cite{Dorland2012, Simms2011, Cheng2009}).  For this reason JMAPS has supported development of a fourth generation detector, the H4RG-10 B1, designed in part to reduce IPC to values that are negligible for astrometry.  

In this paper we discuss the data analysis and modeling that have been used to understand crosstalk for astronomical detectors.  In Section 2 we describe the models for charge diffusion and IPC that were combined to create a $^{55}$Fe model based on the H4RG-10 B0 detector (third generation).   In Section 3 we describe the JPL camera used to collect the data.  In Section 4 we discuss the analysis methodology for single-pixel-reset data and compare the results with the modeled data.  In Section 5 we discuss the analysis methodology used for $^{55}$Fe data obtained with the astrometric camera, and again compare the analysis and modeling results.  Finally, in Section 6 we summarize our findings.   (In a subsequent paper we show how crosstalk affects astrometry and photometry using detailed detector and optics models for a realistic astrometric telescope (See \cite{Dorland2012} for details.))

\section{Crosstalk models}

\subsection{Introduction to charge diffusion and IPC}

Crosstalk between pixels is caused by two independent phenomena, charge diffusion, and IPC.   Charge diffusion is the lateral movement (i.e. pixel-to-pixel) of charge between the points of charge production and charge collection in the bulk substrate of the detector.  Charge diffusion occurs in all photosensitive material, including the silicon substrate of all CCDs.  IPC is the capacitance that arises between adjacent detector pixels in the source-follower CMOS design, and leads to coupling of signal between those pixels via displacement currents flowing from the collection node. 

For optical detectors with photosensitive silicon material, crosstalk can be measured by exposing the detector to a radioactive source with a known charge production rate in bulk silicon, and then measuring the voltage in pixels surrounding a central hit.  $^{55}$Fe is a soft X-ray source, quickly decaying (half-life: 2.7 years) to Mn when a K shell electron is absorbed into the nucleus.   An electron from either the L or M shells drops to fill the hole created in the K shell.  This drop to a lower energy level causes the emission of either a K$_{\alpha}$ (5.9 KeV) or K$_{\beta}$ (6.5 KeV) X-ray.  The absorption of a K$_{\alpha}$ photon in the bulk silicon of the detector produces, on average, 1620 e$^-$, while K$_{\beta}$ photon absorption produces an average of 1778 e$^-$. 

In $^{55}$Fe testing, the detector, under controlled environmental conditions, is briefly exposed to the radioactive source and the resulting charge is collected and read out for each exposure.  When analyzed, photon hits approximately normal to the detector surface and centered on a single pixel are averaged and normalized to create a kernel representative of pixel crosstalk.  $^{55}$Fe testing and analysis is described in more detail in Section 5 of this document.

In addition to $^{55}$Fe testing, single-pixel-reset testing (SPR) allows for the direct characterization of IPC alone, requiring no illumination source [7].   In SPR, after setting all pixels in the SCA to a single voltage and making an initial readout, a well-spaced grid of single pixels is reset to a second voltage level.  A subsequent readout of the pixels will reveal any IPC as signal in pixels adjacent to the reset pixels.  As with $^{55}$Fe testing, multiple SPR test results are averaged (discarding any bad pixels, edge-affected pixels, etc.) to create a representative detector IPC kernel.  SPR testing and data analysis methodology is described in more detail in Section 4.  

The IPC kernel obtained through SPR testing, when combined with charge diffusion modeling, can be used for $^{55}$Fe test verification and prediction.  This resultant $^{55}$Fe model is based on the convolution of a charge diffusion kernel, resulting from an incident K$_{\alpha}$ photon, with an IPC kernel built on SPR-measured IPC.  This new, modeled kernel can be used to approximate the average pixel crosstalk expected in $^{55}$Fe testing for a given detector.  The subsections below describe the modeling of IPC and charge diffusion.

\subsection{IPC model} \label{sec_ipcmodel}

\begin{figure} [t] 
\centerline{\includegraphics{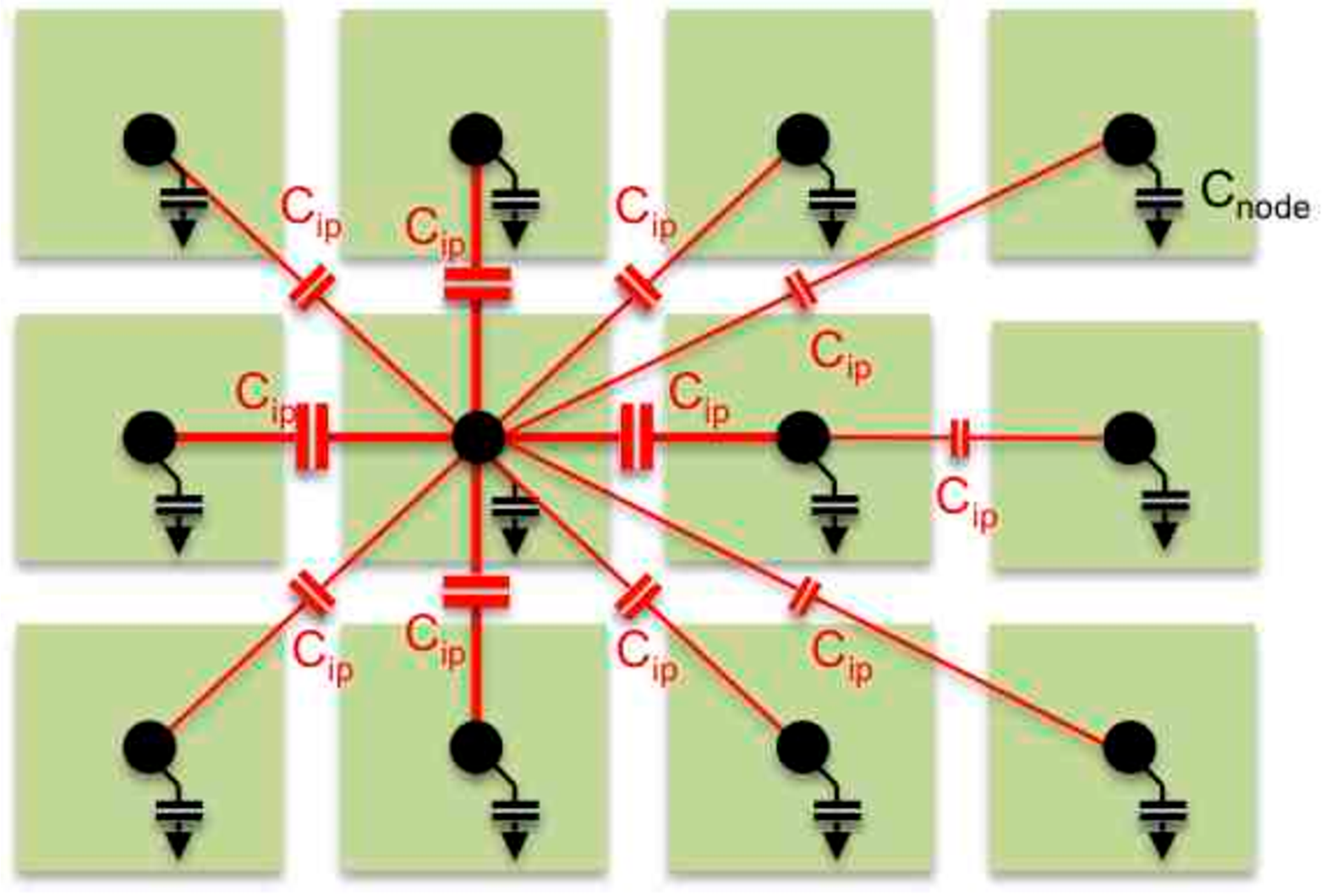}}
\caption{Inter-pixel capacitance (C$_{ip}$).  xtalk00F1.ps}
\label{fig_cip}
\end{figure}

A simple model of the IPC expected within a detector array can be developed based on the approach described by Moore, et al. in 2004 \cite{Moore2004}.   The detector array of photodiodes is modeled as an array of capacitors, as shown in Figure \ref{fig_cip} , each identical and with a capacitance that is unchanging with voltage.

The detector is constructed in a way that allows an electrical field to exist between neighboring collection nodes, essentially creating small coupling capacitors between the nodes. Charge entering a single nodal capacitor $Q_{total}$ causes a voltage change in that node, and through the coupling capacitors, causes voltage changes in neighboring nodes.  $Q_{total}$ is the sum of all the apparent charge seen both in the voltage of the original node and the voltages of $n$ neighboring nodes.

\begin{equation}
\sum_n V_n = \frac{Q_{total}} {C_{node}}
\end{equation}

\begin{figure} [t] 
\centerline{\includegraphics{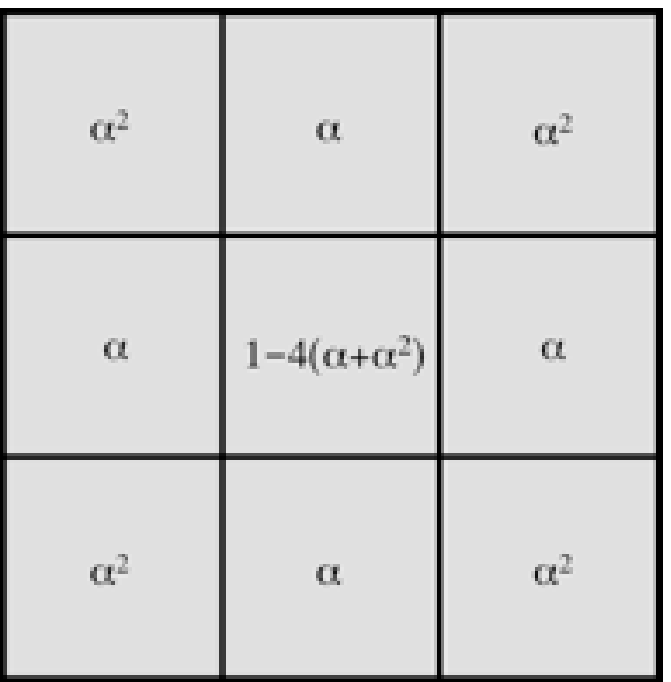}}
\caption{IPC model for 3x3 pixel array.  xtalk00F2.eps}
\label{fig_moore}
\end{figure}

In MooreÕs approach, the impulse response of each node $h(n)$ is a ratio of the charge that appears electrically in a node $Q_n$, to the photocurrent that entered the original node $Q_{total}$,

\begin{equation}
h(n) = \frac{Q_n} {Q_{total}}.
\end{equation}
Using this approach, and assuming two-dimensional symmetry in nearest neighboring pixels and diagonal neighboring pixels, a simple model of IPC can be constructed.  Charge appearing electrically in surrounding pixels is defined in terms of a single variable $\alpha$, defined as the percent of total charge seen in any of the four nearest neighbor pixels.   Symmetry in nearest neighbors and diagonal neighbors is assumed in this model.   An example is shown in Figure \ref{fig_moore} for a 3x3 pixel array.

The sum of the charge in all pixels in the array = 1.0.  The above model kernel, with $\alpha$ = 0.075, is an IPC model for a general CMOS detector.  The model can be extended to larger kernel models to account for detectors with IPC that is more broadly spread.

\subsection{Charge diffusion}

\begin{figure} [t] 
\centerline{\includegraphics{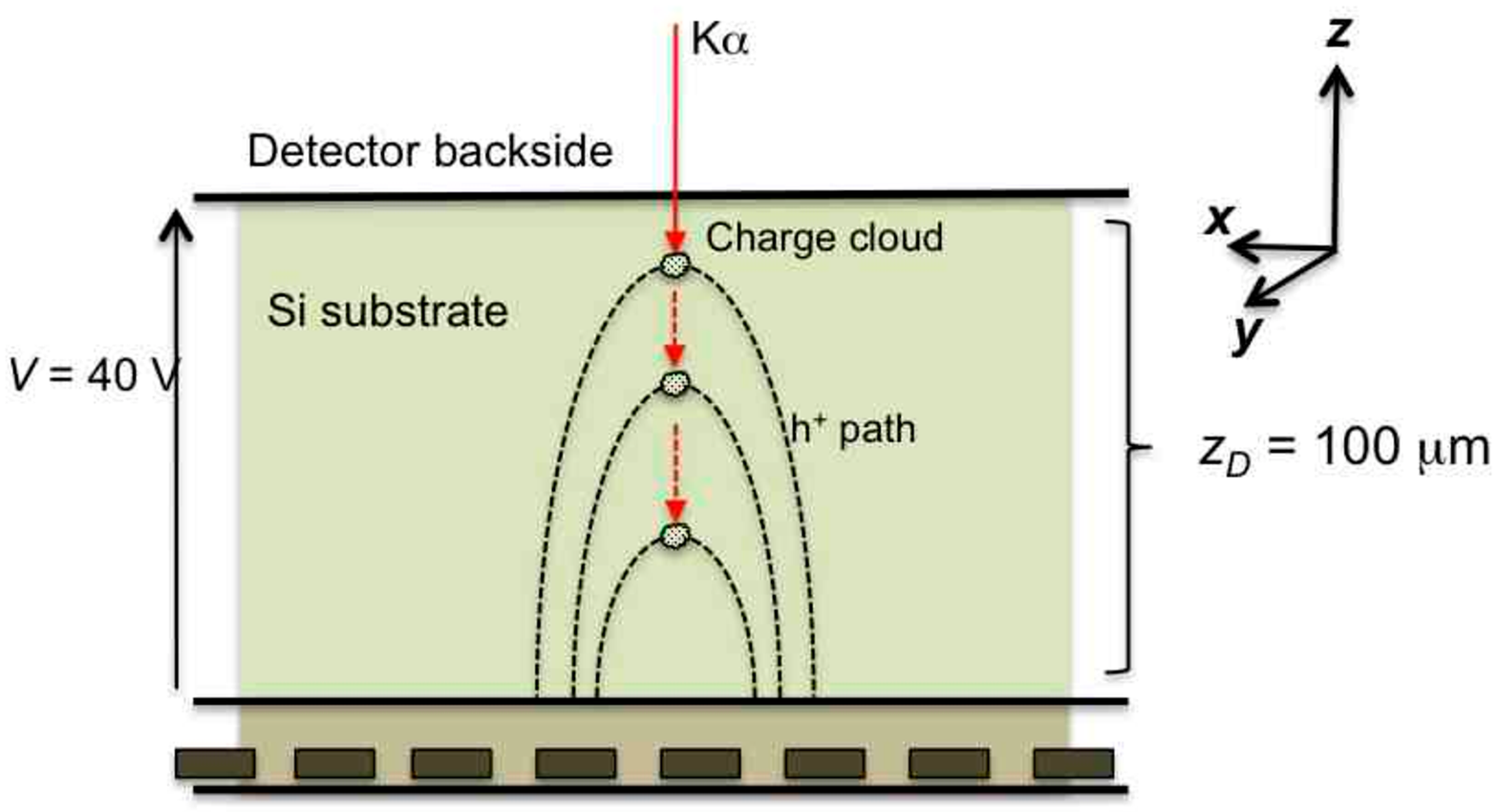}}
\caption{Charge diffusion decreases with increasing absorption depth along $\bf{z}$.  xtalk00F3.ps}
\label{fig_cd}
\end{figure}

Lateral charge diffusion occurs while charge moves between the point of generation and the point of collection, in the detector substrate.  The process begins with the absorption of a K$_{\alpha}$ photon, which, for $^{55}$Fe, produces a ÔcloudÕ of 1620 charge pairs within the substrate, as illustrated in Figure \ref{fig_cd}.

The photon attenuation length $l_a$ for a K$_{\alpha}$ (5.9 KeV) photon is:

\begin{equation}
l_a = \frac{\lambda} {4 \pi \rm{Im}\left( n \right)},
\end{equation}
using the absorptive term Im($n$) of  the complex refractive index of silicon, $n$

\begin{equation}
n = 1 - \delta - i\beta ,
\end{equation}
where $\delta$ is the refractive index decrement and $\beta$ is the absorptive index.  For a photon with $E$=5.9 KeV, $\beta \sim 6.0\times 10^{-7}$ \cite{Henkle} and $l_a\sim28 \mu$m.  In the $^{55}$Fe model, charge diffusion is calculated for thin slices of the bulk Si substrate.  The probability of photon absorption is calculated for each slice:

\begin{equation}
P_{\Delta z} = \frac{1} {l_a}\left(e^{\frac{-d_1}{l_a}}-e^{\frac{-d_2}{l_a}}\right)
\end{equation}
where $d_1$ and $d_2$ are positions along the $\bf{z}$ axis (a direction normal to the pixel array plane ($xy$ plane), as defined for illustration in Figure \ref{fig_cd} above), and $d_1$ is further from the detector.  (In this model, the incident photon is always normal to the detector surface.  This orientation was selected to mimic the symmetry selection criteria for Ôgood hitsÕ used in actual $^{55}$Fe test data analysis, as described below in Section \ref{sec_fedata}.)

We now consider charge diffusion for a detector with a fully depleted substrate.  Charge diffusion, as a function of depth and weighted for absorption probability, is well represented by the two dimensional Gaussian

\begin{equation}
f_{\rm{CD}} = \frac{P_{\Delta z}} {\sqrt{2 \pi \sigma \left(z\right)}}e^{\frac{-x^2 - y^2} {2\sigma\left(z\right)^2}}
\end{equation}
with $\sigma \left(z\right)$ the Root Mean Square (RMS) standard deviation of charge spreading in $x$ and $y$.   This spreading is described in terms of the diffusion constant $D_{\rm{P}}$ and the transit time $t_{\rm{tr}}$ for charges (holes, for the detectors under consideration) to move from the point of charge pair generation to the point of collection

\begin{equation}\label{sig}
\sigma \left( z \right) = \sqrt{2 D_{\rm{P}} t_{\rm{tr}}}.
\end{equation}
The transit time $t_{\rm{tr}}$ can be found using the hole drift velocity $v_{\rm{drift}}$ as described by \cite{Sze1981, Holland1997}:

\begin{equation}
v_{\rm{drift}} = \frac {dz} {dt} = \mu E\left( z \right) = \mu \left( E_{\rm{max}} + \frac {q N_{\rm{d}}} {\epsilon_{\rm{Si}}} z \right),
\end{equation}
and integrating over the entire depletion depth,

\begin{equation}
t_{\rm{tr}} = \frac{\epsilon_{\rm{Si}}} {q N_{\rm{d}}} \ln{\left( \frac{E_{\rm{max}}} {E\left( z \right)}\right) }
\end{equation}
where $\epsilon_{\rm{Si}}$ is the dielectric constant for silicon ($11.9 \times 8.854\times 10^{-12}$ C/V), $q$ is the fundamental charge unit (C), and $N_{\rm{d}}$ is the doping density for the bulk silicon substrate.  $E_{\rm{max}}$ includes the effects of both the substrate bias voltage $V$, and the depletion voltage $V_{\rm{D}}$ \cite{Janesick2001},

\begin{equation}
V_{\rm{D}} = \frac{q N_{\rm{d}}} {2 \epsilon_{\rm{Si}}} z^2_{\rm{D}}
\end{equation}

\begin{equation}
E_{\rm{max}} = - \left( \frac {V} {z_{\rm{D}}} + \frac {q N_{\rm{d}}} {2 \epsilon_{\rm{Si}}} z_{\rm{D}} \right) .
\end{equation}
When V is much larger than $V_{\rm{D}}$ \cite{Holland1997, Janesick2001},

\begin{equation}
\sigma \left( z \right) \approx \sqrt{ \frac{2z^2_{\rm{D}}k_{\rm{B}} T}{qV}}
\end{equation}
where the Einstein relation, $D_{\rm{p}}/ \mu = k_{\rm{B}}T/q$ has been applied to equation \ref{sig}, above, and where $T$ is the temperature of the detector (K) and $k_{\rm{B}}$ is BoltzmannÕs constant (J/K).  For absorption of the photon at a specific $z$ within the substrate:

\begin{equation}
\sigma \left( z \right) = \left[\frac{2 k_{\rm{B}} T \epsilon_{\rm{Si}}} {q^2 N_{\rm{d}}} \ln \left(\frac{E_{\rm{max}}} {\frac {VqN_{\rm{d}}}{z_{\rm{D}}2 \epsilon_{\rm{Si}}} \left(2z - z_{\rm{D}}\right) }\right)\right]^{1/2}.
\end{equation}
In the model, the charge diffusion function $f_{\rm{CD}}$ is calculated for each incremental slice through $z$, using the above equation for $\sigma \left( z \right)$, with:  $T = 193$ K, $V = 40$ V, and $z_{\rm{D}} =100 \mu$m.

\begin{figure} [t] 
\centerline{\includegraphics{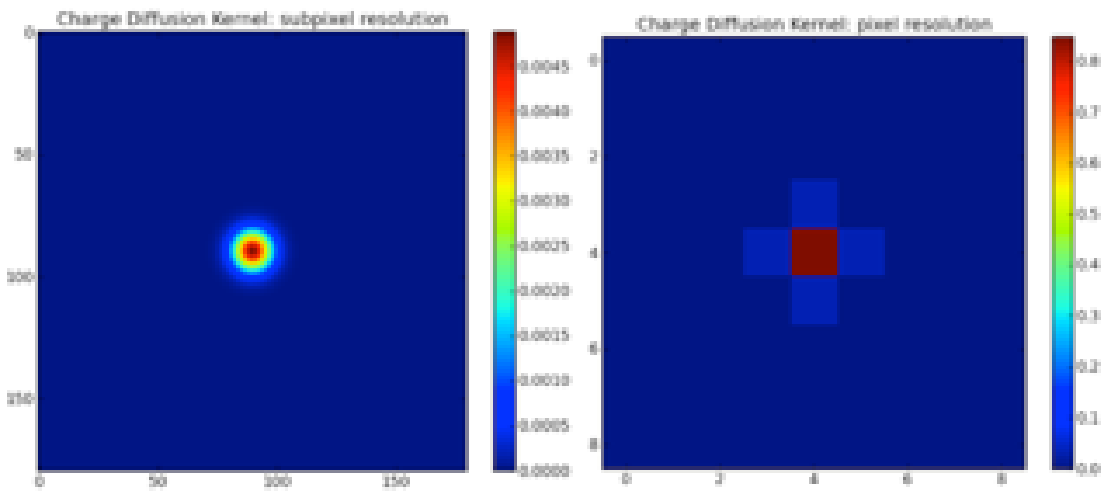}}
\caption{Charge diffusion example images, with $Q_{\rm{pixel}}/Q_{\rm{total}}$ for each pixel.  xtalk00F4.eps}
\label{fig_hilo}
\end{figure}

The charge cloud produced by absorption of a K$_{\alpha}$ photon is resolved at a higher-than-pixel resolution, and diffusion to each neighboring pixel is determined by the position of the cloud center projected somewhere on the surface of the central pixel.  Because of this, the charge diffusion is calculated at a subpixel resolution, with an absorption probability weighting for each slice.  The high-resolution kernel is then summed and rebinned to a lower resolution kernel, ready for convolution with the IPC kernel.  Examples of the charge diffusion model to both subpixel and pixel resolutions are shown in Figure \ref{fig_hilo}.

\section{JMAPS test camera}

\begin{figure} [t] 
\centerline{\includegraphics{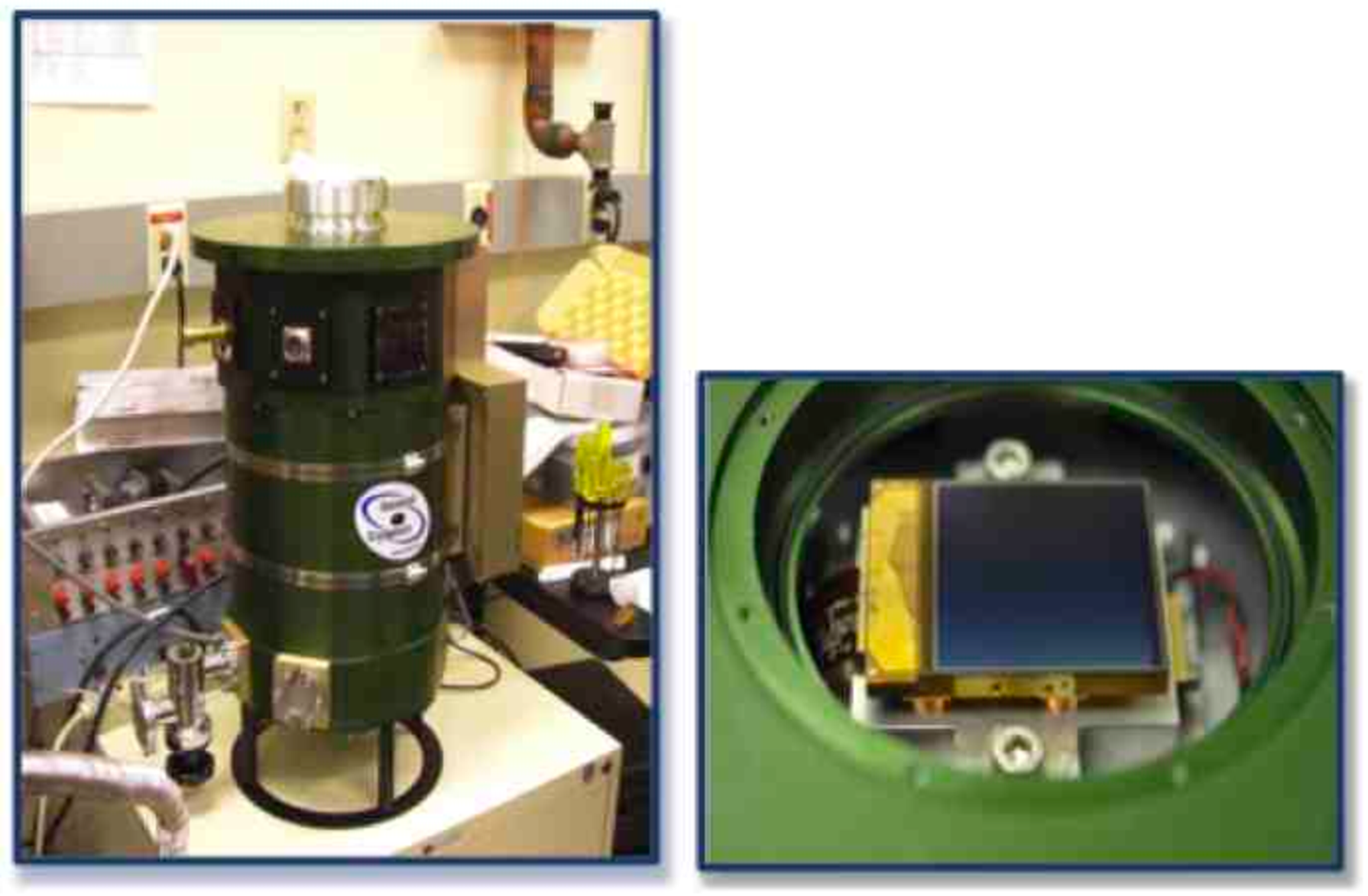}}
\caption{JMAPS camera and H4RH-10 B0 in the dewar.  xtalk00F5.ps}
\label{fig_cam}
\end{figure}

The data described here were taken at the Jet Propulsion Laboratory (JPL) using a test camera specifically designed for ground based astrometric testing of H4RG-10 detectors for JMAPS.  The camera functions with 32 outputs controlled by a TIS non-cryogenic SIDECAR (system image, digitizing, enhancing, controlling, and retrieving) ASIC (Application Specific Integrated Circuit).  The measurements were taken at 193K and the detector substrate voltage was set to 10, 20, 30 or 40V depending on the measurement.  The camera voltage settings were the same for IPC SPR and $^{55}$Fe data acquisition.  Table \ref{tab_cam} lists the primary voltage parameters used for these measurements.  An image of this camera is shown in Figure \ref{fig_cam}.

\begin{table} 
\centering
\caption {\label{tab_cam} JMAPS Camera Voltage Parameters}
\begin{tabular} { l  c  c } \hline
Parameter & Symbol & Voltage (V)\\
\hline \hline
Bias Voltage& VSUB & 10-40\\
Digital Positive Power Supply & VDD & 3.12\\
Analog Positive Power Supply & VDDA & 3.12\\
Drain Node of Pixel SF & CELLDRAIN & 0.00\\
Drain Node of Output SF & DRAIN & 0.00\\
Source Node of Internal Current for SF & VBIASPOWER & 3.00\\
Detector Substrate Voltage & DSUB & 0.50\\
Detector Reset Voltage & VRESET & 0.32\\
Bias Voltage for Current Source of SF & VBIASGATE & 2.00\\
\hline
\end{tabular}
\end{table}

\section{Single pixel reset (SPR) data}

\subsection{SPR data acquisition and analysis}

The JMAPS astrometric test camera is equipped with microcode that can perform the single pixel reset (SPR) technique described by Seshadri \cite{Seshadri2008} and Finger \cite{Finger2006}.  The single pixel reset function enables reset of single pixels in the array to a different reset voltage than the rest of the pixels in the array.  The effective result is charge on the integration node for these reset pixels.  This method is useful for isolating IPC, since the charge is not generated in the photodetector material, and is therefore not susceptible to the effects of charge diffusion.  

\begin{figure} [t] 
\centerline{\includegraphics{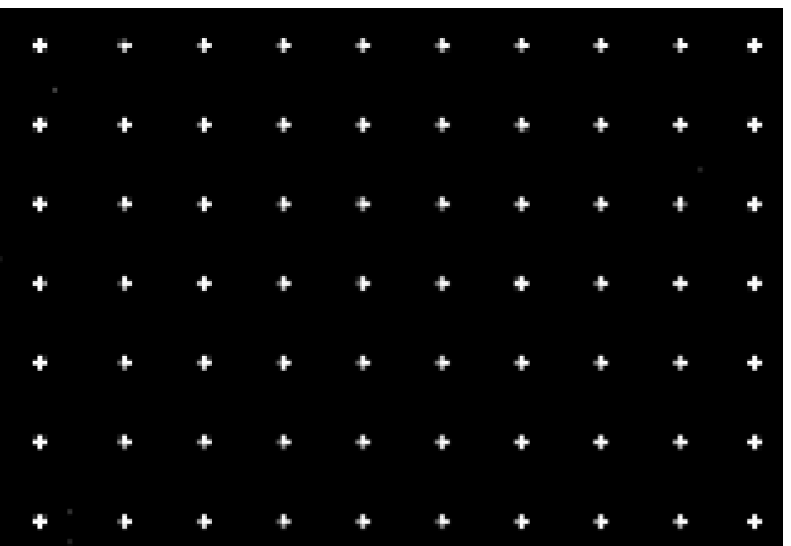}}
\caption{Image of pixels that have been reset using the SPR method for estimating IPC.  xtalk00F6.eps}
\label{fig_spr}
\end{figure}

IPC maps are created using the following data acquisition method.  First the entire detector is reset to a given reset voltage and the pixels are read out to generate an offset frame. Then, a grid of widely spaced single pixels is reset to a new voltage.  Following this reset, all pixels are again read out.  If IPC is present, the difference of these two images will show voltage in pixels neighboring the reset single pixels.  A cropped single pixel reset image of a section of an H4RG-10 B0 detector taken with the JMAPS camera is shown in Figure \ref{fig_spr}.

\begin{figure} [t] 
\centerline{\includegraphics{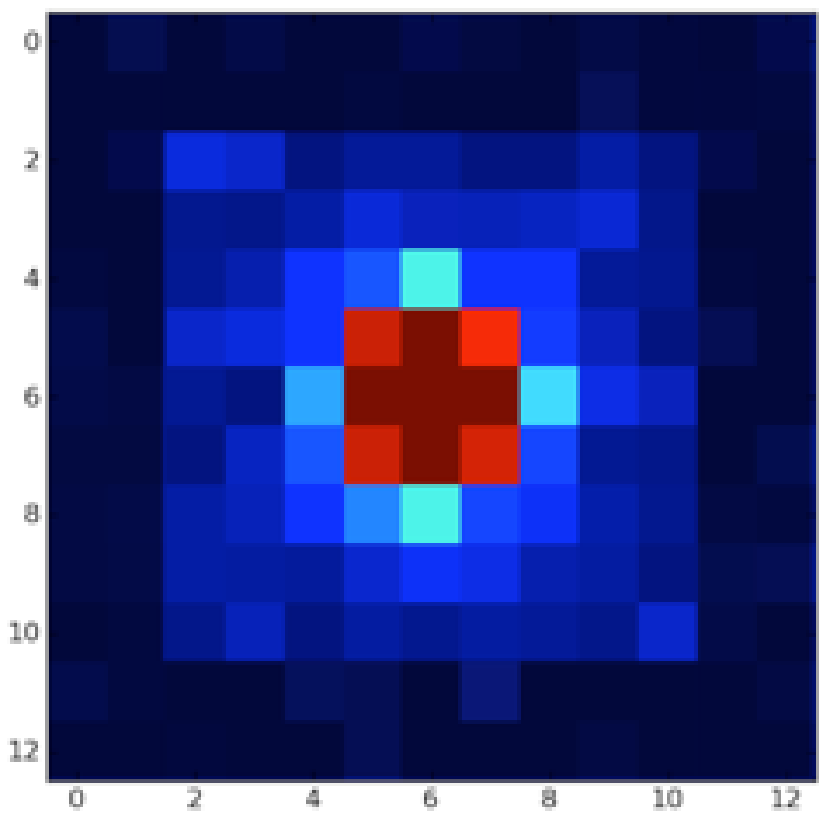}}
\caption{Illustration of background area used for correction.  The shaded region of 2-pixel width is the annulus used for correction.  xtalk00F7.eps}
\label{fig_bc}
\end{figure}

For this analysis the initial reset voltage was first set to 300mV for all pixels.  The SPR grid of pixels was then reset to 0, 100, 200, 300, 400, 500, 600mV.  This resulted in 7 measurements of varying signal levels per substrate voltage (10V, 20V, 30V, 40V).  To analyze the data, the following procedure for finding good hits was adopted:
\begin{enumerate}
\item First the image was offset (bias) corrected.
\item A small range of central pixel values were identified as ÒhitsÓ.  The range of values for the central pixel was identified as a Gaussian distribution in the histogram of the image, and was well above the noise floor of the image. Defining this range is important to ensure that hot or bad pixels are excluded from the IPC estimate.  On average, each image contains approximately 64,000 good hits that are used for the analysis.
\item All pixels defined as ÒhitsÓ based on the criteria above were background corrected using a local background correction method.   An annulus of 2 pixels outside the central hit was extracted. The median of the pixels in this annulus was subtracted from the values of the inner 9x9 pixels.  An image illustrating the annulus and kernel is shown in Figure \ref{fig_bc}.  
\item The background-corrected kernels were then stacked and averaged.  
\item The resultant measured IPC kernel was normalized, so that the charge in a single pixel was represented as a percent of the total charge in the kernel.  
\end{enumerate}
For this detector, the noise floor was reached in both IPC and $^{55}$Fe crosstalk data at the 9x9 pixel boundary.

\subsection{IPC data and model}

\subsubsection{Data}

\begin{figure} [t] 
\centerline{\includegraphics{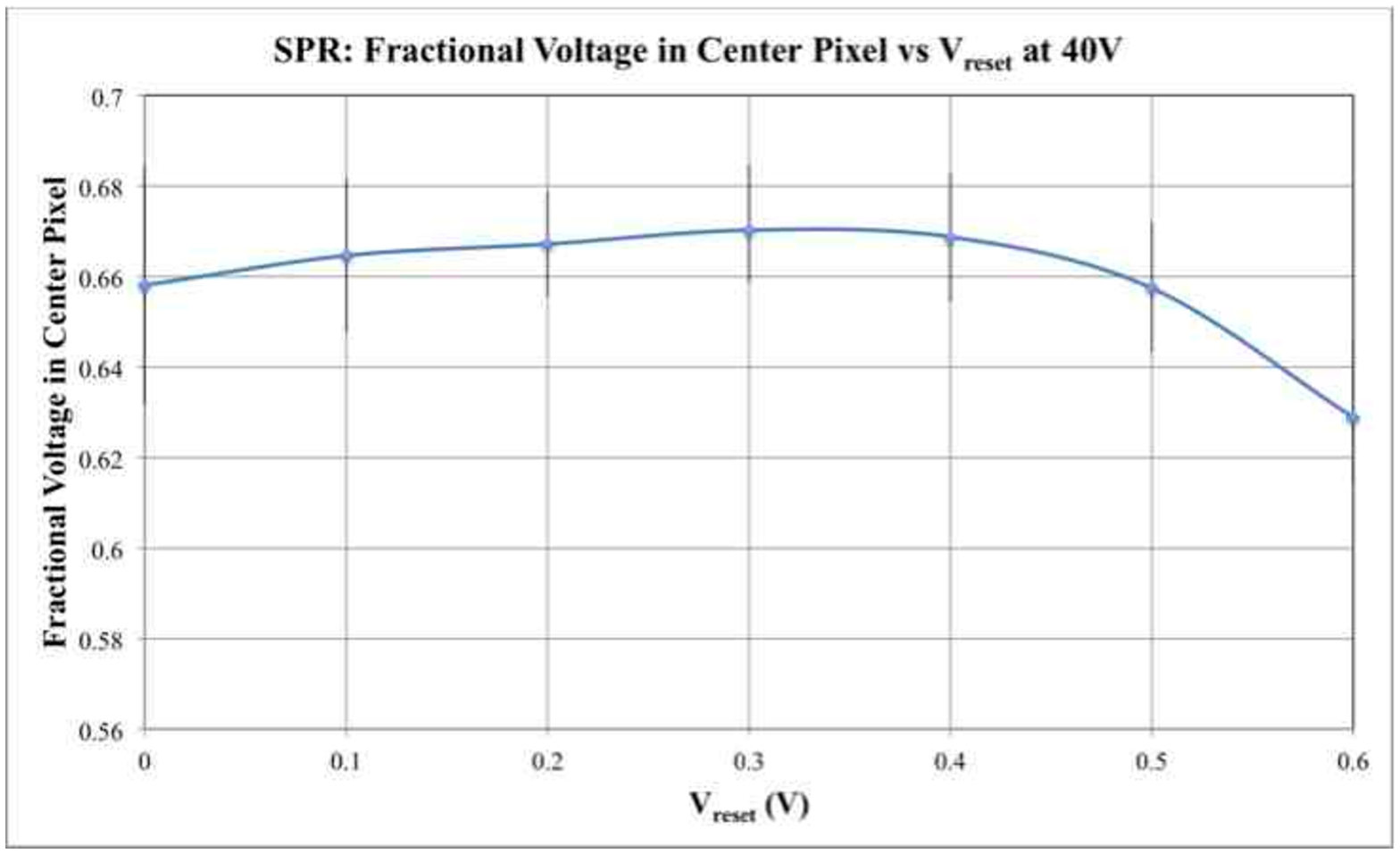}}
\caption{Fractional voltage in the central pixel as a function of the reset voltage level for SPR data.  xtalk00F8.ps}
\label{fig_spr_g1}
\end{figure}

Figure \ref{fig_spr_g1} shows the fraction of signal in the central pixel as a function of the SPR voltage.  In this plot, the saturation limit is indicated by the drop-off in fractional signal at 400 mV. However, in order to be certain that no saturated pixels were included in the final averaged kernel, and to ensure high S/N, the 200 mV SPR dataset was chosen for this analysis (as opposed to 0, 100 and 300 mV SPR data).

Figure \ref{fig_ipc_data} below shows the 9x9 pixel kernel resulting from the data taken at 40V and 200mV SPR.  Despite the fact that the outer 9x9 annulus in the kernel below appears to be statistically indistinguishable for the errors, we note that all values in this annulus are systematically above zero indicating that the noise floor has not been reached.  For this reason, the 9x9 annulus is included in the kernel.   The values listed represent the percentage of the total signal for an average kernel, based on approximately 64,000 events.  Measurement errors represent the standard deviation across all SPR events.

\begin{figure} [t] 
\centerline{\includegraphics{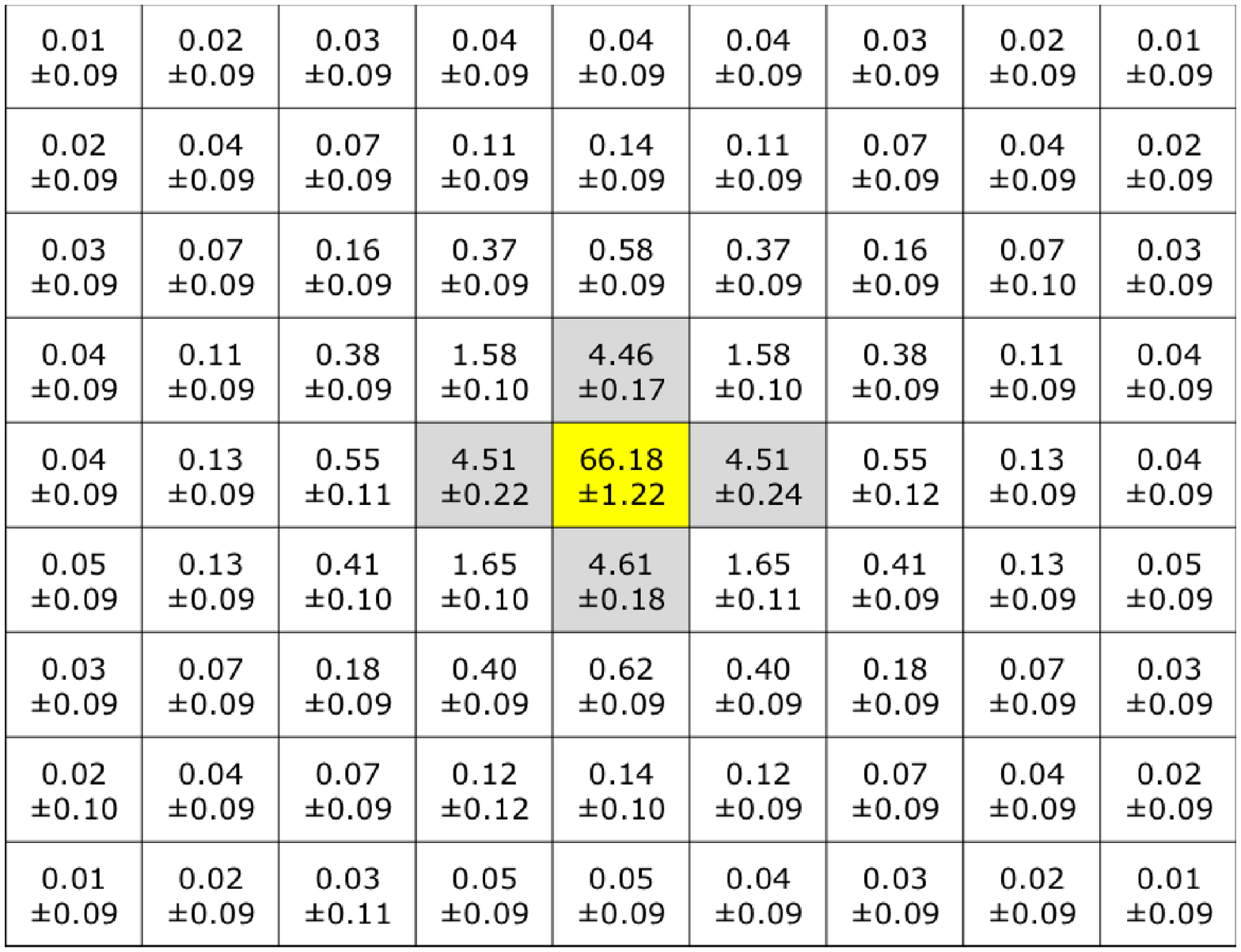}}
\caption{IPC values for H4RG-10 B0 based on single pixel reset method.  xtalk00F9.ps}
\label{fig_ipc_data}
\end{figure}
\clearpage

\subsubsection{Model}

As can be seen in the IPC data kernel in Figure \ref{fig_ipc_data}, IPC can spread significantly beyond the 3x3 grid depicted in Section \ref{sec_ipcmodel}.  In fact, for each detector tested using SPR, the IPC distribution extended to at least a 9x9 grid.  Additionally, each detector type (where type is determined by changing design and processing) was found to have a unique IPC distribution signature.   Based on these findings, the IPC model is an averaged set of SPR data for each detector type, and given in terms of $\alpha$ (the average fractional signal in nearest neighboring pixels, per Section \ref{sec_ipcmodel}).   The use of $\alpha$ is in keeping with the standard modeling approach for IPC and provides a variable to capture the standard deviation of the measured and summed SPR kernels in the $^{55}$Fe model (discussed in Section \ref{fe55dm}).  Variation of $\alpha$ is also needed to model the effects of varying levels of optical crosstalk on centroiding precision.

In the example above (Figure \ref{fig_ipc_data}), $\alpha$ = (4.51 + 4.61 + 4.61 + 4.51)/4.0 = 4.56.   Using this $\alpha$, an IPC model (Figure \ref{fig_ipc_model}) was developed to capture the signal-spreading signature of the B0 generation.  Each cell is a factor of $\alpha$:

\begin{figure} [t] 
\centerline{\includegraphics [width=16 cm] {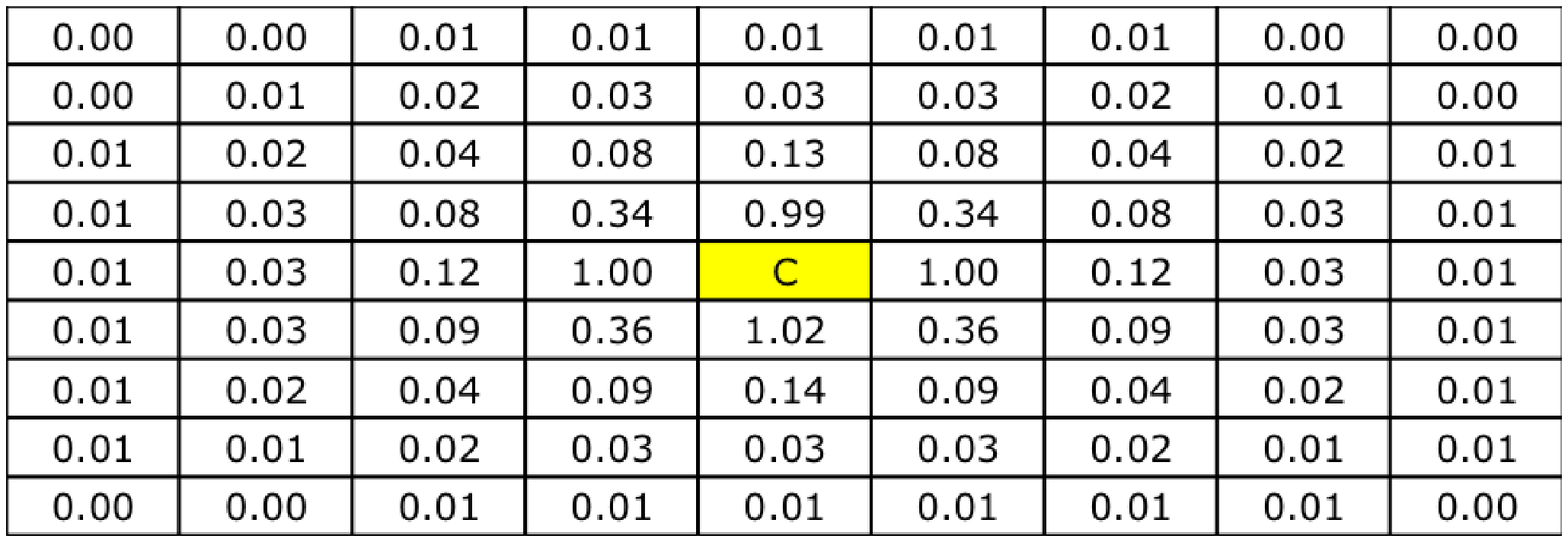}}
\caption{IPC model based on SPR data.  xtalk00F10.ps}
\label{fig_ipc_model}
\end{figure}
where the central pixel, $C = (1/\alpha)  \sum$ (of the grid of coefficients).   



\subsection{IPC asymmetry findings}

\begin{figure} [t] 
\centerline{\includegraphics{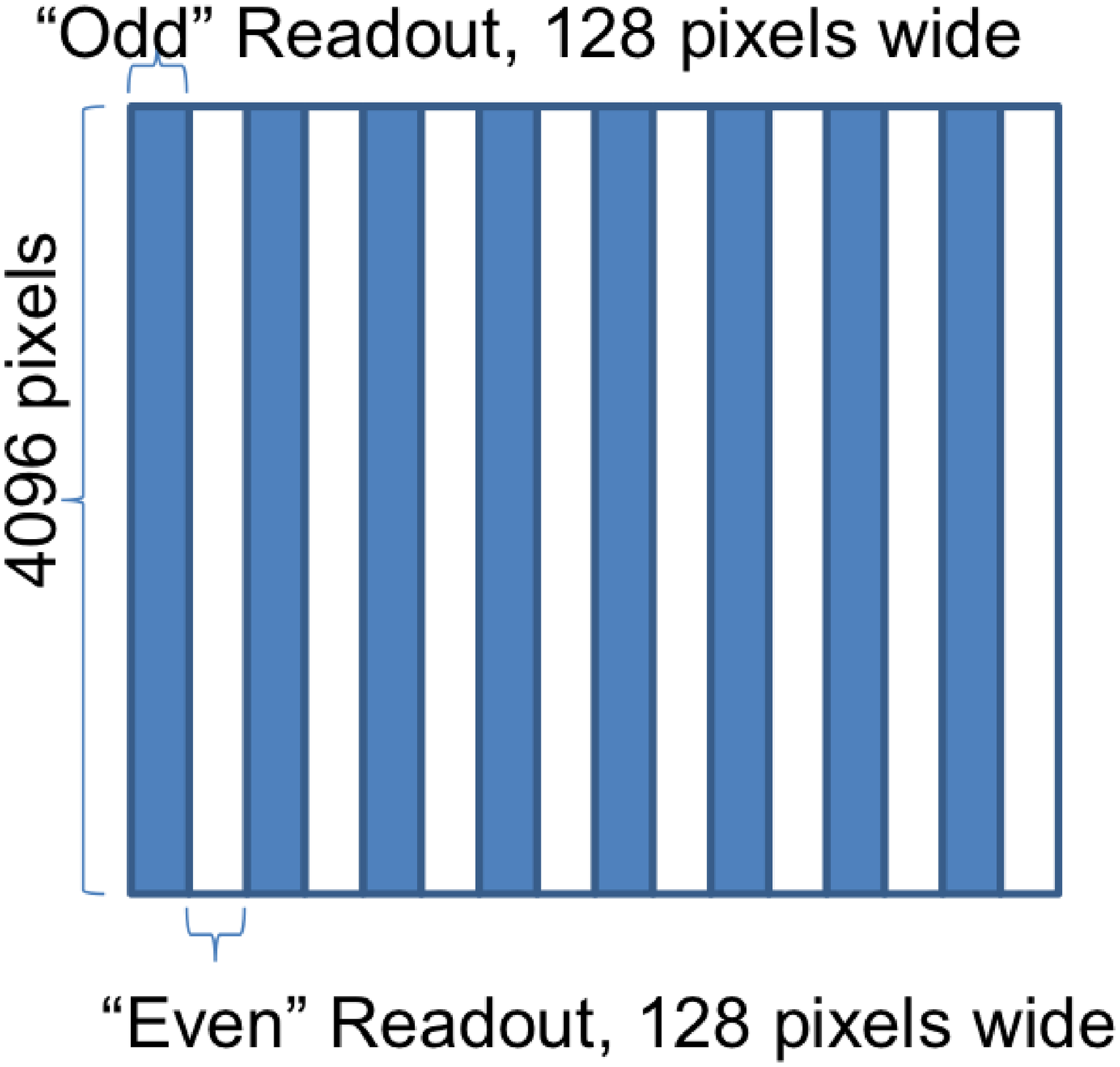}}
\caption{Readout "columns" determined by alternating readout direction.  xtalk00F12.ps}
\label{fig_col}
\end{figure}

Given the ROIC readout asymmetry in these detectors, we specifically set out to determine if a corresponding asymmetry exists in the IPC.  The JMAPS test camera at JPL is designed to read out the full 4096x4096 pixel detector using 32 equally spaced outputs.  This means that the detector is virtually divided into 32, 128x4096 pixel ÔcolumnsÕ.  The columns are read out alternately from the right and left, resulting in all odd columns being read out in one direction and even columns in the other direction.   Figure \ref{fig_col}  shows this left/right, odd/even dichotomy.

\begin{figure} [t] 
\centerline{\includegraphics{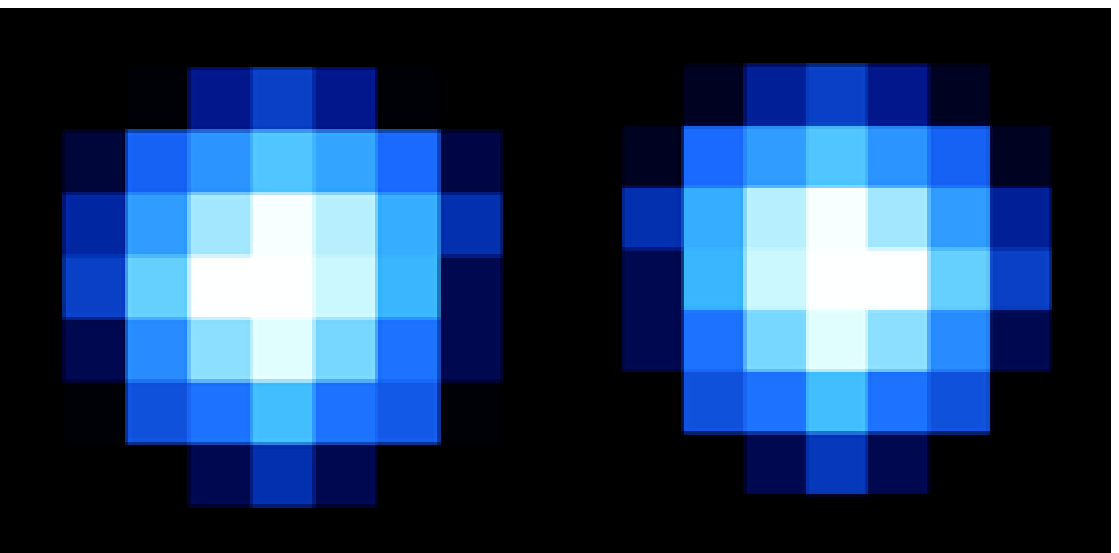}}
\caption{SPR-generated IPC kernels from even (left) and odd (right) columns.  xtalk00F13.eps}
\label{fig_oddeven}
\end{figure}

We conducted a column-by-column comparison of the IPC charge distribution for B0 detectors.  To perform this analysis the hits from all even and all odd columns were separately averaged.  The first and last columns of the array were excluded from this analysis, since they suffer high dark current at the edges that could contaminate the results.  A clear horizontal dichotomy resulting from the direction of the readout is visible in the odd vs even columns, as shown in Figure \ref{fig_oddeven}. 
 
We performed this analysis for all of the SPR data taken for 0-300mV V$_{\rm{reset}}$ values and for all 4 substrate voltages.  Table \ref{tab_asyipc} summarizes our findings.  In the readout direction (horizontal direction) the offset induced by the readout is approximately 5\% depending on whether the column is odd or even.  In addition, a vertical bias is evident in every column (both even and odd) of approximately 3.5\%.  The fifth and eighth columns of Table \ref{tab_asyipc} show the residuals resulting when hits from both odd and even columns are averaged.  We find the residuals to be low in the horizontal/read direction after averaging odd and even columns, but a ~3.5\% residual is still present in the vertical direction.  These findings will be used in the selection criteria of $^{55}$Fe hits, as described in Section \ref{sec_fedata}.

\clearpage

\begin{table} 
\centering
\caption {\label{tab_asyipc} Results of the odd/even column systematic offset test}
\begin{minipage} {13cm}
\centering
\begin{tabular} { c  c || c  c  c  c  c  c } 
\hline
Vsub\footnote{substrate voltage} (V) & Vreset\footnote{SPR reset voltage} (V) & x$_{e}$\footnote{$\rm{x_{right}/x_{left}}$ for all even columns} & x$_{o}$\footnote{$\rm{x_{right}/x_{left}}$ for all odd columns} & x$_{error}$\footnote{\% residual error from hits in both columns} & y$_{e}$\footnote{$\rm{y_{upper}/y_{lower}}$ for even columns} & y$_{o}$\footnote{$\rm{y_{upper}/y_{lower}}$ for odd columns} & y$_{error}$\footnote{\% residual error from hits in both columns}\\
\hline \hline
10.00 & 0.30 & 1.08 & 0.93 & -0.23\% & 0.94 & 0.94 & 6.01\%\\
10.00 & 0.20 & 1.07 & 0.94 & -0.29\% & 0.97 & 0.97 & 3.27\%\\
10.00 & 0.10 & 1.06 & 0.94 & -0.28\% & 0.97 & 0.97 & 3.34\%\\
10.00 & 0.00 & 1.05 & 0.96 & -0.39\% & 0.96 & 0.96 & 3.52\%\\
20.00 & 0.30 & 1.08 & 0.93 & -0.29\% & 0.97 & 0.97 & 2.87\%\\
20.00 & 0.20 & 1.07 & 0.94 & -0.28\% & 0.97 & 0.97 & 3.16\%\\
20.00 & 0.10 & 1.06 & 0.94 & -0.27\% & 0.97 & 0.97 & 3.32\%\\
20.00 & 0.00 & 1.05 & 0.96 & -0.42\% & 0.97 & 0.97 & 3.39\%\\
30.00 & 0.30 & 1.08 & 0.93 & -0.33\% & 0.97 & 0.97 & 2.96\%\\
30.00 & 0.20 & 1.07 & 0.94 & -0.26\% & 0.97 & 0.97 & 3.27\%\\
30.00 & 0.10 & 1.06 & 0.94 & -0.26\% & 0.97 & 0.97 & 3.38\%\\
30.00 & 0.00 & 1.05 & 0.95 & -0.34\% & 0.97 & 0.97 & 3.43\%\\
40.00 & 0.30 & 1.08 & 0.92 & -0.28\% & 0.97 & 0.97 & 3.07\%\\
40.00 & 0.20 & 1.07 & 0.93 & -0.21\% & 0.97 & 0.97 & 3.32\%\\
40.00 & 0.10 & 1.06 & 0.94 & -0.24\% & 0.96 & 0.97 & 3.49\%\\
40.00 & 0.00 & 1.05 & 0.95 & -0.30\% & 0.96 & 0.96 & 3.53\%\\
Averages & & & & -0.29\% & & & 3.46\%\\
\hline
\end{tabular}
\end{minipage}
\end{table}

\subsection{$^{55}$Fe data acquisition and analysis} \label{sec_fedata}

$^{55}$Fe measurements were taken at the same voltage settings as the SPR measurements, using the JMAPS camera.  For these measurements the source was exposed for up to 12 seconds.  During the first two seconds, a mechanical shutter shielded the detector from the source.  This was done to measure the offset from the zero point, due to reset, without contamination from the $^{55}$Fe hits.  In the remaining 10 seconds the shield was removed and the detector was non-destructively read out every 2 seconds, resulting in 5 frames of a progressively increasing population of $^{55}$Fe hits.
 
\begin{figure} [t] 
\centerline{\includegraphics [width=15 cm] {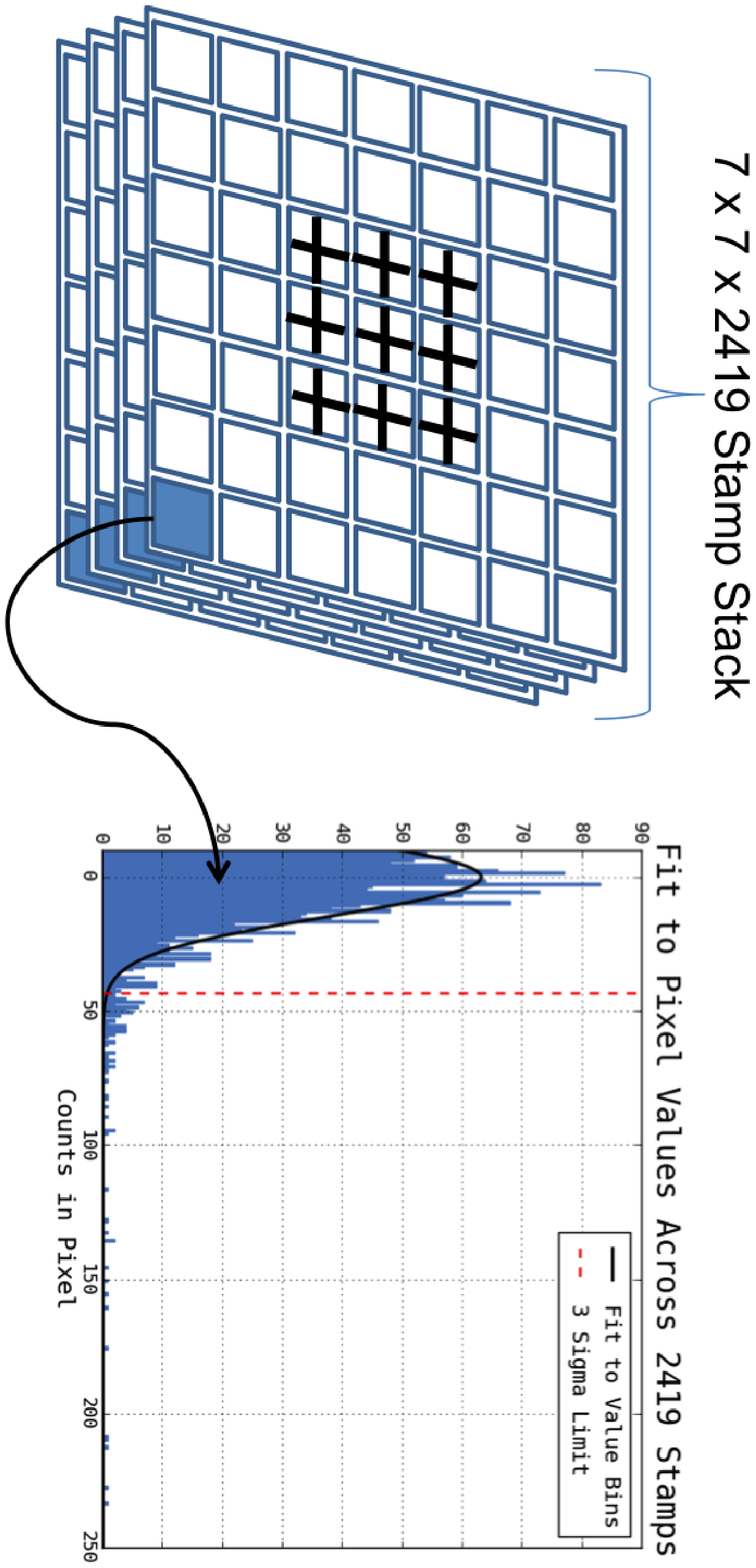}}
\caption{Double hits are removed by fitting a Gaussian to stacked pixels.  Pixels with crosses  were excluded from iteration.  xtalk00F14.ps}
\label{fig_stack}
\end{figure}

	The charge diffusion component of crosstalk is heavily dependent on the depth at which the K$_{\alpha}$ or K$_{\beta}$ photon is absorbed.  Two things must be taken into consideration when selecting good hits.  First, the primary goal is to select hits that are as perpendicular to the detector surface plane as possible.  Secondly, the hits must be selected so that all absorption depths are sampled.  We define an accurate measurement of $^{55}$Fe as the average over all absorption depths.  To ensure this, we have used the following method to define normal $^{55}$Fe hits:

\begin{enumerate}
\item First the image was offset corrected.  To avoid any noise associated with settling, the frames used for this analysis were the fourth and fifth frames.  The fourth frame was taken as the offset frame and subtracted from the fifth frame.
\item A small range of central pixel values were identified as hits.  The range chosen is in ADU and strongly depends on the gain of the system.  Therefore, for this initial rough estimate, the range was taken directly from a subsample of hits on the image without regard to the symmetry of the hit.  The sum in a 9x9 region around the central pixel represents the majority of the analog digital units (ADUs) resulting from the hit.  A Gaussian distribution centered on the K$_{\alpha}$ peak results when these sums are plotted in a histogram.  We defined the median of a Gaussian fit to this distribution divided by the total electrons expected for a K$_{\alpha}$ hit (1620e$^-$) as the gain of the system.  In our case the gain was 1.19 +/- 0.04 ADU per electron.  On average $\sim23,000$ hits proceed to Step 3 as potential K$_{\alpha}$ hits.
\item Double hits or hits falling inside the parameter of the targeted hit region, can skew the crosstalk results.  To remove double hits, all of the potential K$_{\alpha}$ hits from Step 2 were centered and ÔstackedÕ in the $\bf{z}$ direction, a process depicted in Figure \ref{fig_stack}.  Kernels with double hits were removed by fitting a Gaussian distribution along the z-axis of the stack for each pixel except the central 3x3 in the kernel stack.  Hits in the stack with pixel values greater than the mean +/- 3$\sigma$ over the stack were considered double hits and the hit was removed from the analysis.  The remaining hits in the stack are considered Òsingle hitsÓ and were used in the following steps of the analysis. 
\item A local dark correction was then performed on the remaining single hits.  The pixels in an annulus of 2 pixels outside of the 9x9 central hit are defined as background pixels (see Figure \ref{fig_bc}).  The median of the pixels in this annulus was subtracted from the values of the inner 9x9 pixels.
\item The symmetry criteria are perhaps the most challenging criteria to incorporate into the $^{55}$Fe reduction method. Noise and detector effects prevent even a perfectly symmetric hit from appearing symmetric in the data. This leads to confusion of off-center hits with perfectly centered hits contaminated by noise.   To choose perfectly centered hits we use the following criteria: 
\begin{enumerate}
\item The noise of the system can be measured by fitting a Gaussian distribution to all pixels in the image except those affected by $^{55}$Fe hits.  The sigma of this distribution represents the 1-sigma noise of the measurement.  Two times this sigma is the maximum difference in counts that the left/right and upper/lower pixels can have, due to the noise.  For all of our measurements, this noise was $\sim15$ ADU (1$\sigma$). 
\item As discussed in Section 4.3, the B0 detectors show both horizontal and vertical systematic readout asymmetry. The asymmetry in the horizontal direction is column-dependent while the asymmetry in the vertical direction is not. We used the normalized kernels from our IPC analysis to estimate the offset in ADU expected to result from the asymmetry induced by the readout. In most cases this asymmetry in effect contributed $\sim5-10$ADU to the total noise.  

\item With the noise and asymmetry tabulated we defined centered hits as those hits having left/right and upper/lower differences that are less than 2-sigma plus the asymmetry offset.  In most cases this resulted in a total permissible variation between left/right and upper/lower pixels of $\sim35-40$ ADU.
\end{enumerate}

\item Finally, in order to ensure only K$_{\alpha}$ hits are included in the final crosstalk kernel (for the purposes of comparing the data to the model), the sum over all 9x9 pixels of each symmetric hit was taken.  Using the gain from step 2, any hit with a sum from the 9x9 that had values less than or equal to the K$_{\beta}$ escape peak or greater than or equal to the K$_{\beta}$ peak [10] were excluded from the stack.  We note that the peaks of the K$_{\beta}$ lines are used, rather than the edges of these peaks (based on Janesick [10]), in choosing our range.  We do this because the noise is high enough for our system that the K$_{\beta}$ peaks cannot be distinguished from the K$_{\alpha}$ peak when plotted.  To avoid preferentially excluding K$_{\alpha}$ hits that overlap with the K$_{\beta}$ hits due to noise, we established this definition. We recommend using only K$_{\alpha}$ hits for those systems with noise that is low enough to distinguish the three peaks.
\item After applying these criteria, the remaining hits were averaged to generate a 9x9 kernel (in the case of the 40V data) in units of ADU.  The crosstalk kernel was defined as the charge in a single pixel divided by the total charge in the 9x9 kernel. 
\item This test was repeated 9 times, resulting in 9 kernels for all 9 datasets.  The standard deviation across all nine measurements for each pixel represents the error on the final kernel due to statistical differences between each data set. 
\end{enumerate}

This method describes the reduction strategy for the 40V $^{55}$Fe data.  With lower substrate voltage settings we find that a kernel larger than 9x9 is needed to capture all of the escaped charge from the central pixel.  We discuss this in more detail in Section \ref{sec_mav}.

\subsection{$^{55}$Fe 40V data and model} \label{fe55dm}

\subsubsection{Data}

\begin{figure} [t] 
\centerline{\includegraphics{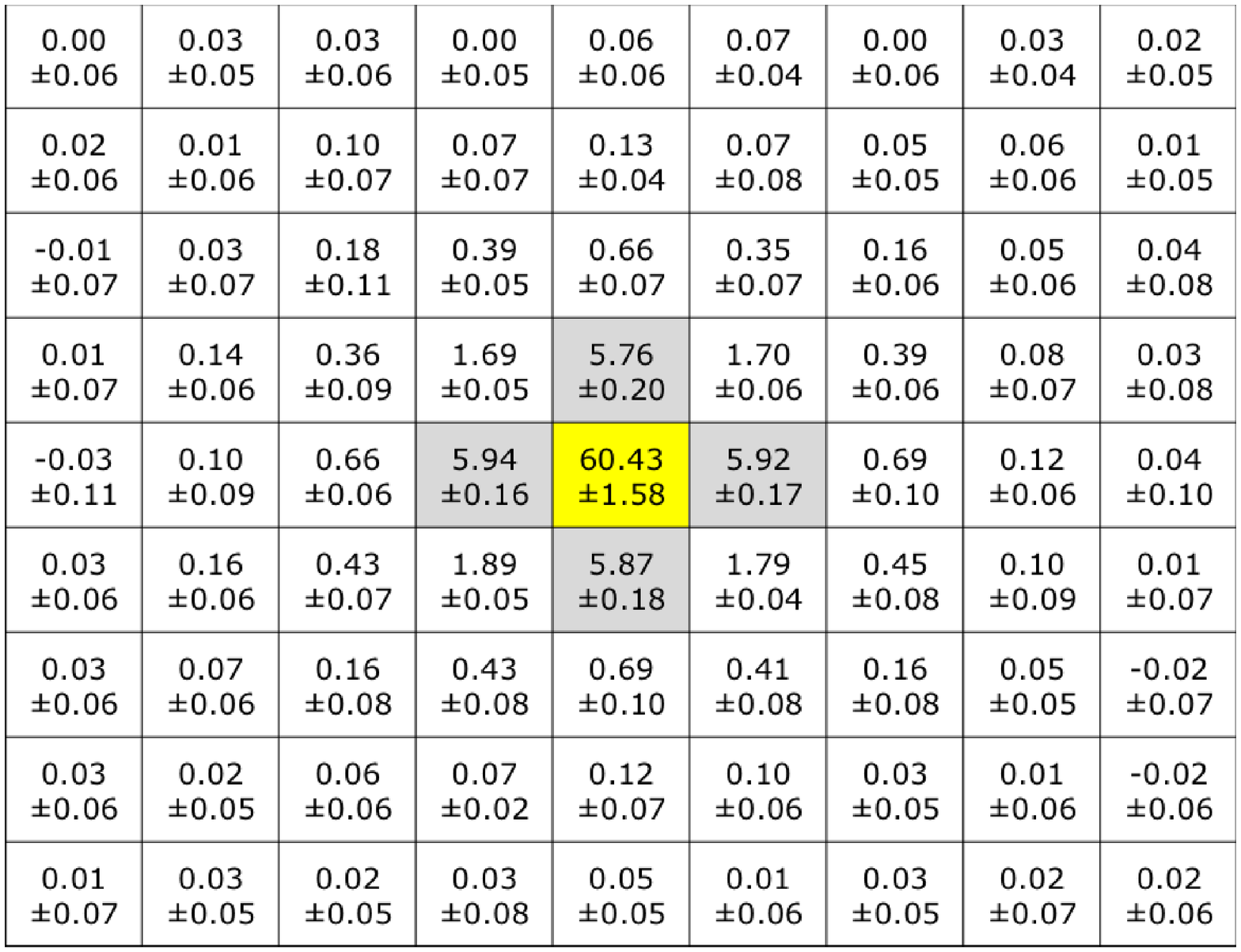}}
\caption{Kernel and errors, with pixel values resulting from $^{55}$Fe tests of H4RG-10 B0.  xtalk00F15.ps}
\label{fig_fedata}
\end{figure}

Figure \ref{fig_fedata} shows the $^{55}$Fe kernel resulting from the acquisition and analysis strategy described in Section \ref{sec_fedata} for the H4RG-10 B0 detector. Kernel pixel values are the percent measured signal for an average of 9 tests.  The total number of Ògood hitsÓ used for this kernel was 1094.  The errors are also shown for each pixel value and represent the standard deviation of the pixel values over all 9 datasets.   We define pixel crosstalk as the value of the average four nearest neighbor pixels (shown in grey) divided by the central value (shown in yellow).  For the H4RG-10 B0 detectors we obtain a measured crosstalk value of 9.83\% for this $^{55}$Fe data set.

\subsubsection{Model}

\begin{figure} [t] 
\centerline{\includegraphics{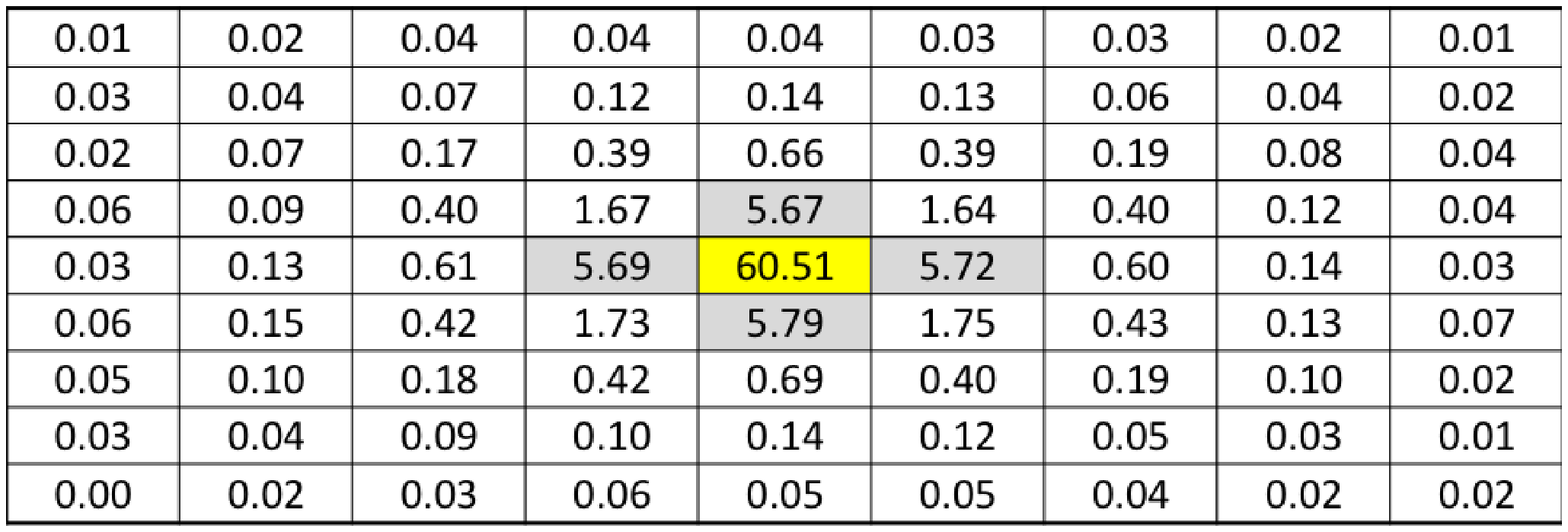}}
\caption{Predicted kernel, with pixel values resulting from the $^{55}$Fe model.  xtalk00F16.ps}
\label{fig_femodel}
\end{figure}

To generate the $^{55}$Fe model, the average IPC kernel ($f_{\rm{IPC}, ij}$) , as described above in Section 4.2, is used to represent the IPC spreading typical of the detector.  The uncertainty associated with the standard deviation ($\sigma_{\rm{cp}}$) of the averaged single-pixel-reset (SPR) results is included in the $^{55}$Fe model through convolution of the charge diffusion kernel with a set of IPC kernels derived from new alphas ($\alpha_{\rm{new}}$).  These $\alpha_{\rm{new}}$ values are calculated from a set of central pixel values ($C_{\rm{new}}$) selected from a Gaussian distribution built around the averaged central pixel value ($C_{\rm{ave}}$) and the standard deviation of that averaged value, $\sigma_{\rm{cp}}$:

\begin{equation}
C_{\rm{new}} = \sigma_{\rm{cp}} \left[-2 \ln{\left(1 - rn_1\right)}\right]^{1/2}\cos{\left(2 \pi rn_2 \right)} + C_{\rm{ave}}
\end{equation}

\begin{equation}
\alpha_{\rm{new}} = \frac{\left(1 - C_{\rm{new}} \right)} {\sum_{ij}f_{\rm{IPC}, ij}}
\end{equation}
where, $rn_1$ and $rn_2$ are random numbers in the interval (0, 1) (uniform deviates).  The convolution is performed for a suitably large number of $\alpha_{\rm{new}}$ values, with noise added to each convolved $^{55}$Fe kernel.  (Noise values are randomly selected from a Gaussian distribution, with a noise standard deviation ($\sigma_{\rm{noise}}$),

\begin{equation}
\sigma_{\rm{noise}} = \frac {\epsilon_q} {1620 G}
\end{equation}
where $G$ is the measured gain, $\epsilon_q$ is the expected background noise (in e$^-$), and 1620 is the number of charge pairs generated by a K$_\alpha$ photon absorbed in silicon.)  The $^{55}$Fe kernels are summed, and the final set of pixels is normalized so that the sum of charge across all pixels in the final kernel = 1.  The two-dimensional convolution of the charge diffusion and IPC kernels is performed using the Python routine convolve2d.py.  Figure \ref{fig_femodel} shows the results of the $^{55}$Fe model.

	  A comparison of this $^{55}$Fe model kernel with the measured $^{55}$Fe data in Figure \ref{fig_fedata} indicates that the combination our SPR model and charge diffusion model predicts the actual $^{55}$Fe kernel very well at 40V.  The central values for the model kernel are within the errors of the actual $^{55}$Fe data as both tables above show.   The SPR and charge diffusion model also predicts the values in the surrounding pixels very well.  We note that some of the pixels around the central pixel do not quite match the model to within the errors.  This is likely because the signal in the surrounding pixels is much less than the central pixel.  When these pixels are normalized to the total in the kernel, the effect of noise in the normalization is much higher for these pixels than the central pixel.  Thus, in addition to the statistical errors across all hits, there is error in the normalization due to the noise of the system.  A system with lower noise than the camera presented here will likely find model values even closer to the observed values.  We explore the predictive power of our model in more detail in the next section.

\subsection{$^{55}$Fe modeling across voltages} \label{sec_mav}
	
	$^{55}$Fe measurements were taken at substrate voltages ranging from 10 to 40V.  This section describes the crosstalk results for all voltage settings and compares these results with the model predictions. 
	 
When extracting the kernel for the 10, 20, and 30 V measurements, it is critical to take into consideration broader spreading induced by charge diffusion.  For 10, 20 and 30 V settings, we found that 15x15, 13x13, and 11x11 square pixel kernels respectively were needed to capture all of the charge from the $^{55}$Fe hit.  Because the gain of the system should not change with substrate voltage, the gain from the 40V measurement was used for the lower voltage settings, since here the K$_{\alpha}$ distribution is much tighter. 

	Table \ref{tab_fedm} shows the central and neighboring pixels resulting from the $^{55}$Fe measurements and those predicted by the model.  	

\begin{figure} [b] 
\centerline{\includegraphics {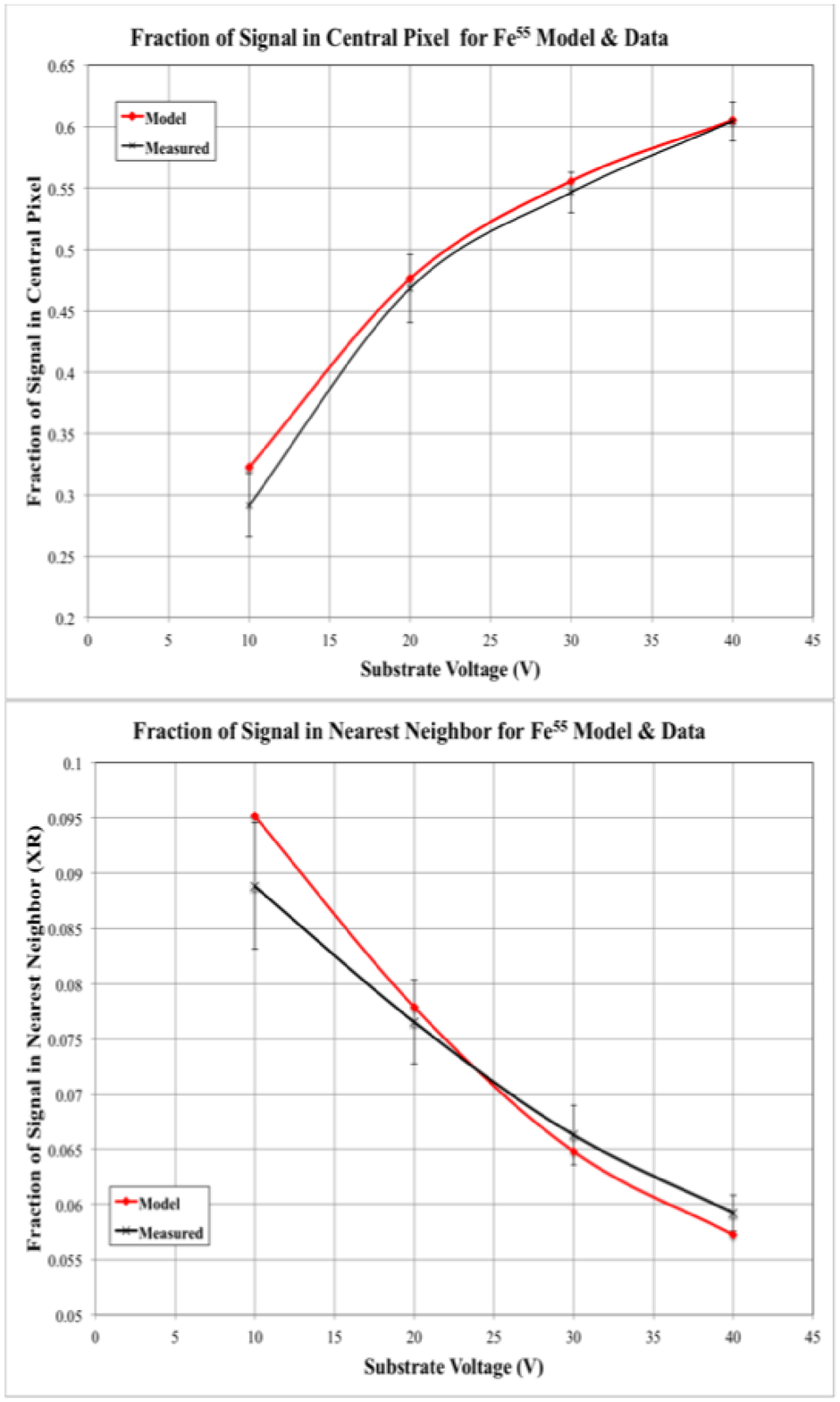}}
\caption{Comparison of central  and nearest neighbor (right and left) pixel values: predicted by the model, and taken from the final $^{55}$Fe kernel.  xtalk00F17.ps}
\label{fig_feg1}
\end{figure}

\begin{table} 
\centering
\caption {\label{tab_fedm} $^{55}$Fe model and data results for substrate voltages 10-40V}
\begin{tabular} { c  c  c  c  c } \hline
\multicolumn {5} { c } {Model and Data Comparison for 10-40V $^{55}$Fe}\\ 
\hline \hline
 & & \multicolumn {3} {c } {Central Pixel}\\
 \hline
Vsub(V) & kernel size & model & data & std dev\\
\hline
10 & 15x15 & 32.24\% & 29.16\% & 2.61\%\\
20 & 13x13 & 47.63\% & 46.85\% & 2.79\%\\
30 & 11x11 & 55.57\% & 54.65\% & 1.65\%\\
40 & 9x9 & 60.51\% & 60.43\% & 1.58\%\\
\hline
 & & \multicolumn {3} {c } {X-right Pixel}\\
 \hline
10 & 15x15 & 9.52\% & 8.88\% & 0.57\%\\
20 & 13x13 & 7.79\% & 7.65\% & 0.38\%\\
30 & 11x11 & 6.48\% & 6.63\% & 0.27\%\\
40 & 9x9 & 5.72\% & 5.92\% & 0.17\%\\
\hline
 & & \multicolumn {3} {c } {X-left Pixel}\\
 \hline
10 & 15x15 & 9.50\% & 8.88\% & 0.25\%\\
20 & 13x13 & 7.77\% & 7.67\% & 0.41\%\\
30 & 11x11 & 6.48\% & 6.57\% & 0.36\%\\
40 & 9x9 & 5.71\% & 5.94\% & 0.16\%\\
\hline
\end{tabular}
\end{table}
\clearpage
Figure \ref{fig_feg1}  contains plots of the central pixel value and nearest neighbor value, both for the measurements and for the model. While the model comes close to the results we have for the 10V data, it does not predict the data to within the errors.  At 10V the distributed charge inhabits a very large kernel:  15x15 pixels, for which the charge is very close to the noise floor of the system.  We therefore conclude that the 10V data presented here is not high fidelity.

\clearpage


\section{Summary}

	While the design of source-follower CMOS-hybrid arrays eliminates some noise sources such as charge transfer inefficiency (CTI), read noise and dark current, it also results in an increase in IPC, in contrast to other detectors.  IPC can be problematic for some astronomical applications like astrometry and photometry, which relies on high S/N to obtain accurate and scientifically useful results. 
	
	 In this paper we have used the SPR method to model IPC in TIS H4RG-10 arrays.  We combined a charge diffusion model with this IPC model to understand the global effect of crosstalk on these arrays.  Our primary findings are as follows:
\begin{itemize}
\item The IPC in the H4RG-10 B0 detectors extends out to 9x9 pixels based on our single pixel reset method, requiring a extension to MooreÕs IPC model to capture all of the charge spread. 
\item The single pixel reset method shows a read out bias that is column dependent.  This effect introduces an inherent systematic bias in the IPC pattern.  This bias must be included in the $^{55}$Fe reduction strategy in order to extract all good hits from a $^{55}$Fe image.
\item We have developed a model for combining IPC and charge diffusion effects as a function of substrate voltage.  Our model results closely match our empirical results.
\item Our results suggest that the SPR method alone combined with the model of charge diffusion described here can provide enough information to simulate the dominant properties of H4RG-10 B0 crosstalk to high accuracy without need for a radioactive source.  In addition the SPR data with this charge diffusion model can be used to very accurately predict astrometric and photometric performance [4].
\item We find that the H4RG-10 B0 generation of detectors shows crosstalk of $\sim10\%$, and a signal loss of $\sim40\%$ in the central pixel at 40V.    A risk reduction effort is currently underway to reduce this crosstalk by half in a new generation (B1) of H4RG-10 detectors. 
\end{itemize} 
	The model based on SPR data presented here reproduces the $^{55}$Fe data to within the statistical errors and noise properties of our system.   In a companion paper we use the models to explore the effects crosstalk at various levels (5-15\%) have on astrometric and photometric applications [4].
	
\section{Acknowledgements}

The research described in this paper was performed in part by the Jet Propulsion Laboratory, California Institute of Technology, under contract with the National Aeronautics and Space Administration.  The authors are very grateful to Chris Paine, Allan Runkle, Stuart Shaklan, and Larry Hovland, for all of their help designing, building and operating the JMAPS H4RG-10 astrometric camera at JPL.  We are also very grateful to Andre Wong, Bill Tennant, and Yibin Bai at TIS for their informative thoughts and suggestions that have greatly improved this paper.

\end{document}